\documentclass[superscriptaddress,floatfix,preprintnumbers, nofootinbib,hyperref]{revtex4-2} 

\pdfoutput=1
\usepackage[title]{appendix}
\usepackage[colorlinks=true,breaklinks=true]{hyperref}
\usepackage[normalem]{ulem}
\usepackage[utf8]{inputenc}
\hypersetup{allcolors=[rgb]{0.0 0.0 0.6},linkcolor=[rgb]{0.75 0.05 0.05}}
\usepackage{amsmath,amssymb}
\usepackage{epsfig}  
\usepackage{graphicx}   
\usepackage{slashed}       
\usepackage{url}
\usepackage[dvipsnames]{xcolor}
\usepackage{color}
\usepackage{multirow}
\usepackage{comment}
\usepackage{amssymb}
\usepackage[version=3]{mhchem}
\usepackage{orcidlink}

\DeclareMathOperator{\GeV}{GeV}
\DeclareMathOperator{\eV}{eV}

\DeclareMathOperator{\keV}{keV}
\DeclareMathOperator{\MeV}{MeV}

\DeclareMathOperator{\kpc}{kpc}

\DeclareUnicodeCharacter{2212}{-}

\allowdisplaybreaks

\bibpunct{[}{]}{,}{n}{}{,}
\definecolor{darkspringgreen}{rgb}{0.09, 0.45, 0.27}

\begin{document}

\title{Probing axion-like particles with multimessenger observations of neutron
star mergers}

\author{Francesca Lecce~\orcidlink{0009-0009-6967-5042}}
\email{francesca.lecce@ba.infn.it}
\affiliation{Dipartimento Interateneo di Fisica  ``Michelangelo Merlin'', Via Amendola 173, 70126 Bari, Italy}
\affiliation{Istituto Nazionale di Fisica Nucleare - Sezione di Bari, Via Orabona 4, 70126 Bari, Italy}%

\author{Alessandro Lella~\orcidlink{0000-0002-3266-3154}}
\email{alessandro.lella@ba.infn.it}
\affiliation{Dipartimento Interateneo di Fisica  ``Michelangelo Merlin'', Via Amendola 173, 70126 Bari, Italy}
\affiliation{Istituto Nazionale di Fisica Nucleare - Sezione di Bari, Via Orabona 4, 70126 Bari, Italy}%

\author{Giuseppe  Lucente~\orcidlink{0000-0002-3266-3154}}
\email{giuseppe.lucente@ba.infn.it}
\affiliation{SLAC National Accelerator Laboratory, 2575 Sand Hill Rd, Menlo Park, CA 94025}

\author{Vimal Vijayan~\orcidlink{0000-0002-4690-2515}}
\email{v.vijayan@gsi.de}
\affiliation{GSI Helmholtzzentrum f\"ur Schwerionenforschung, Planckstra{\ss}e 1, 64291 Darmstadt, Germany}

\author{Andreas Bauswein~\orcidlink{0000-0001-6798-3572}}
\email{a.bauswein@gsi.de}
\affiliation{GSI Helmholtzzentrum f\"ur Schwerionenforschung, Planckstra{\ss}e 1, 64291 Darmstadt, Germany}

\author{Maurizio Giannotti~\orcidlink{0000-0001-9823-6262}}
\email{mgiannotti@unizar.es}
\affiliation{Centro de Astropartículas y Física de Altas Energías, University of Zaragoza, Zaragoza, 50009, Aragón, Spain}
\affiliation{Physical Sciences, Barry University, 11300 NE 2nd Ave., Miami Shores, FL 33161, USA}

\author{Alessandro Mirizzi~\orcidlink{0000-0002-5382-3786}}
\email{alessandro.mirizzi@ba.infn.it}
\affiliation{Dipartimento Interateneo di Fisica  ``Michelangelo Merlin'', Via Amendola 173, 70126 Bari, Italy}
\affiliation{Istituto Nazionale di Fisica Nucleare - Sezione di Bari, Via Orabona 4, 70126 Bari, Italy}%

\begin{abstract}
Axion-like particles (ALPs) can be copiously produced
in binary neutron star (BNS) mergers through nucleon-nucleon bremsstrahlung if the ALP-nucleon couplings $g_{a N}$ are sizable.  
The ALP-photon coupling $g_{a\gamma}$ may trigger conversions of
ultralight ALPs into photons in the magnetic fields of the merger remnant and of the Milky Way. This effect would lead to a potentially observable short gamma-ray signal,  in coincidence with the gravitational-wave signal produced during the merging process.
This event could be detected through multi-messenger observation of BNS mergers employing the synergy between gravitational-wave detectors and gamma-ray telescopes.
Here, we study the sensitivity of current and proposed MeV gamma-ray experiments to detect such a signal.
As an explicit example, we consider ALP couplings related as in the Kim–Shifman–Vainshtein–Zakharov (KSVZ) axion model, and show that in this case
the proposed instruments  can reach a sensitivity down
to $g_{a\gamma}\gtrsim \textrm{few} \times 10^{-13}\,\text{GeV}^{-1}$ for
$m_a \lesssim 10^{-9}$~eV, comparable with the SN 1987A limit.
\end{abstract}

\date{\today}
\smallskip

\maketitle

\section{Introduction}

Axion-like particles (ALPs) 
are predicted in several extensions of the Standard Model (SM) of particle physics~\cite{Jaeckel:2010ni,Ringwald:2014vqa,DiLuzio:2020wdo,Agrawal:2021dbo,Giannotti:2022euq,Antel:2023hkf}. 
In particular, string theory can account for an ``axiverse'' with the QCD axion~\cite{Peccei:1977hh,Peccei:1977ur,Weinberg:1977ma,Wilczek:1977pj} and several ultralight ALPs~\cite{Arvanitaki:2009fg,Cicoli:2012sz,Cicoli:2023opf}.
 Stellar environments represent ideal engines to produce ALPs, via their interactions with SM particles.
Typically, one expects that ALPs are produced by  (quasi-)thermal 
 processes, so that the stellar ALP spectrum would show typical energies set by the temperature of the stellar plasma (see Ref.~\cite{Carenza:2024ehj} for a review of different stellar axion fluxes). Remarkably, core-collapse supernovae~(SNe), reaching a core temperature $T\sim30$~MeV, are the most energetic sources of stellar ALPs.
Their production from the hot and dense nuclear medium in the SN core can be copious if ALPs would be coupled to nucleons~\cite{Carena:1988kr,Brinkmann:1988vi,Raffelt:1993ix,Raffelt:1996wa,Carenza:2019pxu,Lella:2022uwi,Lella:2023bfb}. The ALP-nucleon interactions are commonly described by the following Lagrangian~\cite{DiLuzio:2020wdo}:
\begin{equation}
{\mathcal L}_{aN}=
\frac{\partial_\mu a}{2m_N}\sum_{N=p,n}
g_{aN}{\bar N}\gamma^\mu\gamma_5 N \,\ ,
\label{eq:nucllagr}
\end{equation}
where $a$ is the ALP field, $m_N$ is the nucleon mass, and $g_{aN}$ is the ALP coupling to nucleon species, with $N=p,n$ for protons and neutrons, respectively. This Lagrangian allows for ALP production via nucleon-nucleon (NN) bremsstrahlung~\cite{Carenza:2019pxu}. An enhancement in ALP emissivity can be achieved if a significant fraction of thermal pions is present in the SN core, allowing for ALP production via pionic Compton-like ($\pi N$) processes~\cite{Carenza:2020cis}. In such a situation, SN 1987A neutrino observations would exclude values of $g_{aN} \gtrsim 10^{-9}$ in order to avoid a significant shortening of the neutrino burst~\cite{Lella:2023bfb}.

Furthermore, one can consider scenarios in which ALPs are also coupled to photons through the following Lagrangian~\cite{Raffelt:1987yu,Raffelt:1987im}:
\begin{equation}
    \mathcal{L}_{a\gamma} = -\frac{1}{4} g_{a\gamma} a F_{\mu\nu} \tilde{F}^{\mu\nu}= g_{a\gamma} a \, \mathbf{E} \cdot \mathbf{B} \, ,
    \label{eq:gag}
\end{equation} 
where $g_{a\gamma}$ is the ALP-photon coupling, $F_{\mu\nu}$ is the electromagnetic field, and  $\tilde{F}_{\mu\nu}=\frac{1}{2}\epsilon_{\mu\nu\rho\sigma}F^{\rho\sigma}$ is its dual. 
In the presence of the ALP-photon coupling, 
ultralight ALPs produced thanks to  the  ALP-nucleon coupling in Galactic SNe ($m_a \lesssim 10^{-10}$~eV) would efficiently  convert within the Milky Way magnetic field, giving rise to a gamma-ray signal
observable by gamma-ray experiments operating in the MeV energy range.
The search of gamma-rays induced by ALPs from SNe would benefit from the simultaneous detection of SN neutrinos, which could serve as an external time trigger, linking the ALP signal to a specific supernova event.

 This connection has been pointed out since the SN 1987A explosion~\cite{Kamiokande-II:1987idp,Hirata:1988ad,Bionta:1987qt,IMB:1988suc,Alekseev:1988gp}.
 At that time, the non-observation of the ALP-induced gamma-ray signal in the Gamma-Ray Spectrometer (GRS) of the Solar Maximum Mission (SMM) in coincidence with the neutrino signal from SN 1987A
provided strong bounds on ALPs from SNe~\cite{Grifols:1996id,Brockway:1996yr,Payez:2014xsa,Hoof:2022xbe,Calore:2020tjw}.
Recently, it has been shown that previous bounds can be strengthened significantly if one also accounts for ALP-photon conversions in the magnetic field of the progenitor star~\cite{Manzari:2024jns}.

Binary neutron star (BNS) merger events represent an environment with similar physical properties with respect to SNe. Therefore,
\emph{mutatis mutandis},
one expects ALP production in the merger remnant by ALP-nucleon coupling {(see Ref.~\cite{Dietrich2019} for merger simulations considering the effects of ALP bremsstrahlung)} and conversions in gamma-rays in the magnetic field of the Milky Way as well as of the BNS remnant itself, thanks to the ALP-photon coupling.
{The  $NN$ bremsstrahlung  process was considered  in Ref.~\cite{Fiorillo:2022piv} as a production process for ALPs in BNS mergers. However, corrections beyond the one-pion-exchange approximation~\cite{Carenza:2019pxu} were not included in that paper. Therefore, the authors obtained an overproduction of ALPs with respect to the correct calculation (see, e.g., Refs.~\cite{Carenza:2019pxu,Lella:2022uwi,Lella:2023bfb} for the SN case).
Furthermore, the ALP production from BNS mergers was also estimated in the recent Ref.~\cite{Manzari:2024jns}. However, in that work, the authors adopted an SN model, rather than a self-consistent model for the BNS merger event, to get a rough estimation of the ALP production from mergers. It is apparent that, at the moment, the characterization of ALP-induced signals from BNS mergers has not been carried at the same level of sophistication as in SNe. We devote this current work to fill this gap.}

Unlike the case of SNe, BNS mergers are likely extra-galactic events.
This prevents the possibility of using the neutrino signal as an external trigger for the ALP-induced gamma-ray burst with current detectors, as is possible in the case of Galactic SNe.
However, for BNS merger events, one expects to detect a gravitational-wave (GW) signal with current detectors like LIGO~\cite{Harry:2010zz}
and VIRGO~\cite{VIRGO:2014yos}; see, e.g., the recent multi-messenger observation of the BNS merger event GW170817~\cite{LIGOScientific:2017ync,Kocevski:2017liw,LIGOScientific:2017zic,Savchenko:2017ffs}.
 
Due to the simultaneity of the ALP-induced gamma-ray burst and the GW signal, one would replace the neutrinos with a GW detection as an external trigger to determine the time at which one has to search for the ALP-induced gamma-ray signal in a gamma-ray telescope.
 Therefore, the possible multi-messenger detection of BNS merger events with GW interferometers plus gamma-ray telescopes widens the search for ALP-induced signals to an extragalactic horizon. 
We  remark that in the case of the 
multi-messenger observation of the BNS merger GW 170817,
\emph{Fermi}-LAT was entering the South Atlantic Anomaly at the time of the LIGO/Virgo trigger  and therefore cannot place constraints on the existence of a gamma-ray burst emission associated with the moment
of binary coalescence~\cite{Kocevski:2017liw}.
However, a gamma-ray burst peaked at energies $E\sim270\,\keV$ was observed by the \emph{Fermi} Gamma-ray Burst Monitor~\cite{LIGOScientific:2017ync} with a delay of 1.7 s with respect to the GW trigger. 
Based on this observation, we assume that the standard gamma-ray burst expected after the merging can be distinguished from the ALP-induced signal due to this 
time delay. This expectation needs further investigations when better theoretical modeling of gamma-ray bursts from BNS mergers are available. Furthermore, it is not clear if every BNS merger produces a beamed relativistic outflow. At any rate, one would not expect a gamma-ray burst signal for the majority of merger events observed from 
equatorial directions.

In the following we discuss this proposal in detail. In Sec.~\ref{sec:NSM model} we present the BNS merger model used in our work. 
Then, in Sec.~\ref{sec:prod} we calculate the ALP energy spectrum from { $NN$ bremsstrahlung}. 
{ In Sec.~\ref{sec:megnetic} we characterize the ALP-photon conversions in the BNS remnant and in  Milky Way magnetic fields}. In Sec.~\ref{ref:gammaray} we briefly describe the gamma-ray telescopes we use in our analysis and present their sensitivity to the ALP-photon coupling $g_{a\gamma}$. Finally, in Sec.~\ref{sec:conclu} we discuss our results and conclude.

\section{BNS merger model}
\label{sec:NSM model}


The merging of compact objects is a complex dynamical process, which requires expensive numerical simulations involving three-dimensional relativistic (magneto-)hydrodynamics, neutrino transport, and a model for the hot nuclear equation of state~(see, e.g., Ref.~\cite{Baiotti:2016qnr}). These simulations show that during the post-merger evolution densities of a few times the nuclear saturation density and temperatures of several tens of MeV occur in the merger remnant.
The merging is an inherently three-dimensional process with a distinguished axis 
defined by the orbital angular momentum of the binary. 
The post-merger remnant evolves into an approximately axi-symmetric object, which, due to the large amount of angular momentum, is strongly deformed. 
As an initial assessment, we consider radially-averaged properties of the merger remnant, which simplify the analysis while still providing transparent and reasonably accurate physical insights.

For the present analysis, we adopt data from a simulation of a 1.375-1.375~$M_\odot$ BNS merger simulation using the DD2 equation of state~\cite{Typel2010}. The calculation was conducted with a three-dimensional relativistic smooth particle hydrodynamics code, which includes a neutrino treatment to model energy losses by neutrinos and neutrino reabsorption~\cite{Ardevol2019,Collins2023}. The simulation covers the post-merger evolution until about 23 ms after merging.

The properties of the characteristic environment produced in the merger are shown in
Fig.~\ref{fig:plothydro}, {where}  we display the radial temperature profile $T$ (upper left panel) and the matter density $\rho$ profile (upper right panel) at different times after merging. 
{The quantities are averaged over polar and azimuthal angles.} 
For $r \lesssim$ 5~km, the temperature declines from $\sim 25$~MeV at $t=5$~ms to $\sim 15$~MeV at
$t=15-20$~ms. At larger distances, the $T$ profile is similar for the different post-merging times considered, and monotonically drops from $T\sim 15$~MeV at $r\sim 10$~km to a few MeV at $r\gtrsim 50$~km. 
The matter density $\rho$  increases with time for $r\lesssim 5$~km, with a maximum in the center from 
$\rho \sim 6\times 10^{14}$~g/cm$^3$
at $t=5$~ms to $7\times 10^{14}$~g/cm$^3$ at $t=20$~ms. At larger distances, the density shows a sharp drop around $r \sim 10 \, \text{km}$, marking the radius of the remnant of the BNS merger.

\subsection{Nuclear matter effects}

\begin{figure}[t]
\centering
    \includegraphics[width=1\textwidth]{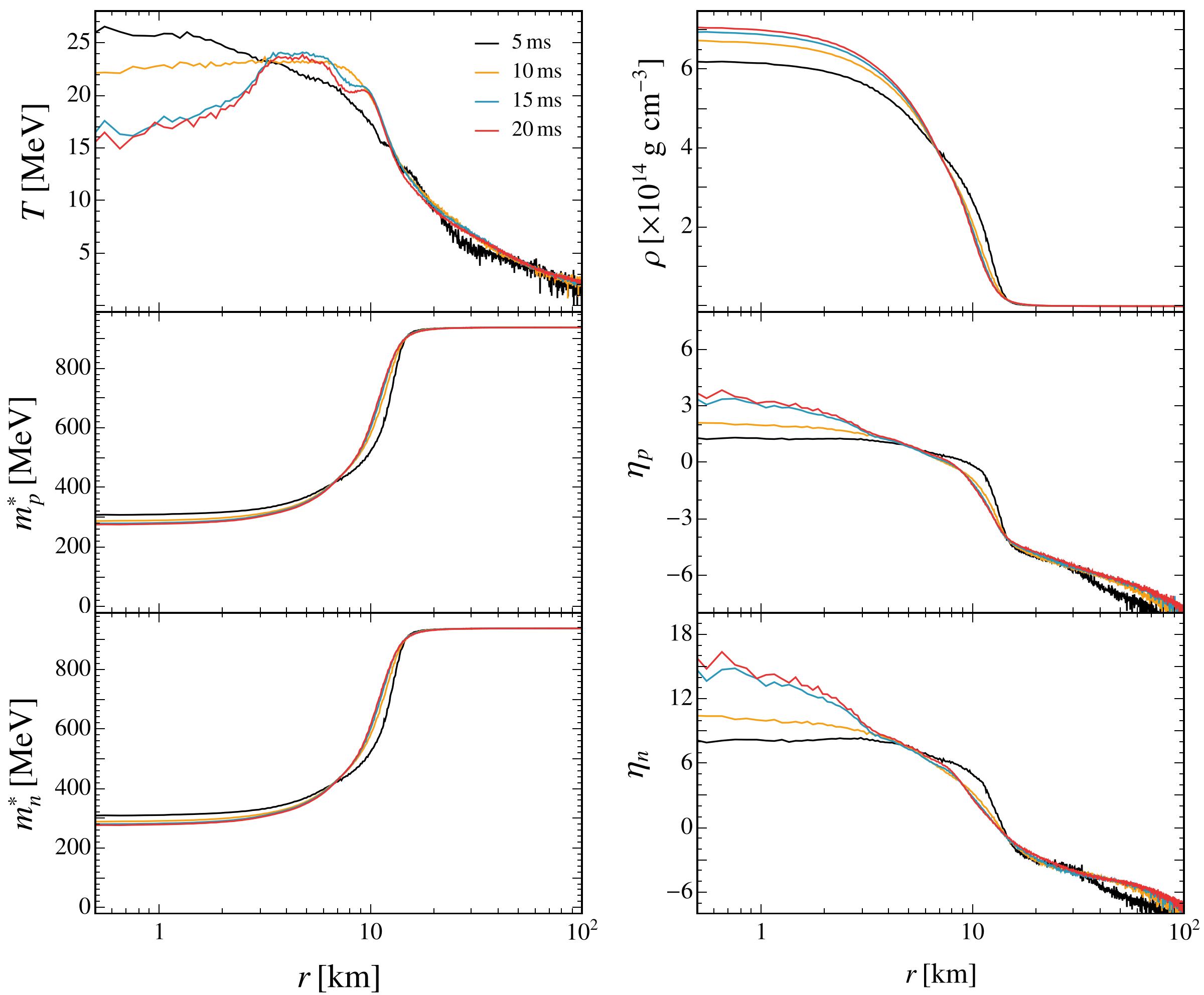}
    \caption{ {Radial evolution at different post-merging times for different quantities from the adopted BNS merger model. \emph{Left panels:} From top to bottom,   temperature $T$, effective proton mass $m_{p}^{*}$ and effective neutron mass $m_{n}^{*}$. \emph{Right panels:} From top to bottom,  density $\rho$, proton degeneracy factor $\eta_{p}$ and  neutron degeneracy factor $\eta_{n}$.
    } 
    }
    \label{fig:plothydro}
\end{figure}

{ In presence of ALP-nucleon Lagrangian in Eq.~\eqref{eq:nucllagr}, ALPs can be produced in BNS mergers by $NN$ bremsstrahlung, involving protons and neutrons as relevant targets~(see Sec.~\ref{sec:prod}).}
To accurately evaluate the ALP production rate, it is crucial to account for key aspects of the nuclear matter present in the BNS merger.
In particular, the very high density reached in these environments, which exceeds nuclear saturation, requires a careful treatment of { nucleon} properties, as standard descriptions may no longer be valid.
The combination of high density and relatively low temperature results in significant degeneracy effects, which cannot be neglected when modeling the { nucleon} population.
Additionally, finite-density effects modify the bare nucleon mass, further influencing their role in ALP production. 
{Accounting for these effects,} the { nucleon} distribution can be written as
\begin{equation}
   f_{N} = \frac{1}{1 + \exp \left[{(E_{N}^{\text{kin}}(p,m_{N}^{\ast}) - \mu^\ast_{N})}/{T}\right]} \, ,
\end{equation} 
where the kinetic  energy is given by the modified dispersion relation~\cite{Iwamoto:1984ir,Hempel:2014ssa,Martinez-Pinedo:2012eaj}
\begin{equation}
  E_{N}^{\text{kin}} = \sqrt{p^{2} + {m_{N}^{\ast}}^{2}},
\end{equation}
where $p$ is the nucleon momentum and $m_{N}^{\ast}$ denotes the \emph{nucleon effective mass}~\cite{Raffelt:1996wa}. This latter is defined as
\begin{equation} 
  m_{N}^{\ast} = m_{N} +   \Sigma_{S}, 
\end{equation}
where $\Sigma_{S}$ is the nucleon scalar self-energy and $m_{p}=938\,\text{MeV}$ ($m_{n}=939\,\text{MeV}$)
is the bare mass of the proton (neutron).
In Fig.~\ref{fig:plothydro} we show 
the radial evolution of 
$m_p^{\ast}$ (middle left panel) and $m_n^{\ast}$ (bottom left panel). 
For $r\lesssim 5$~km, the effective masses of both species can be reduced down to $\sim 300$~MeV.


It is convenient to also consider the  nucleon degeneracy parameters $\eta_{N}$, defined as~\cite{Hempel:2014ssa}
\begin{equation}
    \eta_{N}=\frac{\mu_{N}-m_{N}-U}{T}\, ,
\end{equation}
where $\mu_N$ is the nucleon chemical potential
and $U=\Sigma_{S}+\Sigma_{V}$ the non relativistic mean-field potential of nucleons, {with $\Sigma_{V}$ being the nucleon vector self-energy and $\mu^*_N = \mu_N - U$ the so-called effective or kinetic chemical potential~\cite{Hempel:2014ssa}.}  
In particular, {nucleons} can be considered  essentially degenerate for $\eta_{N} \gtrsim 1$ and non-degenerate for $\eta_{N} < 0$.
Fig.~\ref{fig:plothydro} displays the radial evolution of $\eta_p$ (middle right panel) and $\eta_n$ (bottom right panel) at different times after merging.
One realizes that protons are degenerate for $r \lesssim 6\,\text{km}$ and the transition to a non-degenerate state occurs at ${r \gtrsim 6\,\text{km}}$. Additionally, the degeneracy parameter decreases more rapidly as time increases and presents a sharp drop at ${r \sim 15\,\text{km}}$. 
{On the other hand, neutrons show a significantly higher degeneracy parameter than protons in the inner regions of the remnant, so that they can be considered degenerate until  ${r \lesssim 10\,\text{km}}$ as the transition to a non-degenerate state occurs at $r \gtrsim 10\,\text{km}$.}

\section{ALP production in BNS merger remnant}
\label{sec:prod}

The ALP production via the ALP-nucleon Lagrangian of Eq.~\eqref{eq:nucllagr}   has been widely studied in the context of SNe, where $NN$  bremsstrahlung $N+N\to N+N+a$~\cite{Carena:1988kr,Brinkmann:1988vi,Raffelt:1993ix,Raffelt:1996wa,Carenza:2019pxu}  and  pionic Compton processes 
$\pi+N \to N+a$~\cite{Raffelt:1993ix,Keil:1996ju,Carenza:2020cis} have been considered. In particular,  
recent $NN$ bremsstrahlung calculations have introduced corrections to the naive one-pion-exchange approximation. Notably, these  include many-body effects on the nucleon dispersion relations in the medium and its finite lifetime due to multiple scattering~\cite{Carenza:2019pxu}~(see also Ref.~\cite{Springmann:2024mjp} for a recent work on a self-consistent approach for ALP production in dense environments).
Furthermore, in the presence of a significant fraction of thermal pions,
it has been shown that 
pionic processes may significantly increase the ALP emissivity, as shown in Ref.~\cite{Lella:2022uwi}.  
We refer to this latter paper for the state-of-the-art calculation of the SN emissivities for these processes. 

In the following, we apply the results of these previous works to evaluate the ALP emissivity from BNS due to the ALP-nucleon coupling $g_{aN}$. In particular, we focus on the $NN$ bremsstrahlung process only, neglecting the pionic process. This choice is motivated by the very high densities reached during the merging event, which may lead to the bosonic condensation of pions in vast regions of the environment under study with only a smaller contribution of thermal pions~\cite{Vijayan:2023qrt,Pajkos2025}. The role of pions in ALP production deserves a dedicated study, which we postpone to a future work. Moreover, under these assumptions, our results should be considered as conservative.

Effects due to the strong gravitational field in the BNS, such as time dilation and energy redshift, significantly impact the ALP emission. Here, these effects have been incorporated into the ALP spectrum by introducing the lapse factor $\alpha_{\rm GR}(r)\leq 1$ (see Refs.~\cite{Calore:2021hhn,Lella:2022uwi}).
\begin{figure}
    \centering
    \includegraphics[width=0.7\linewidth]{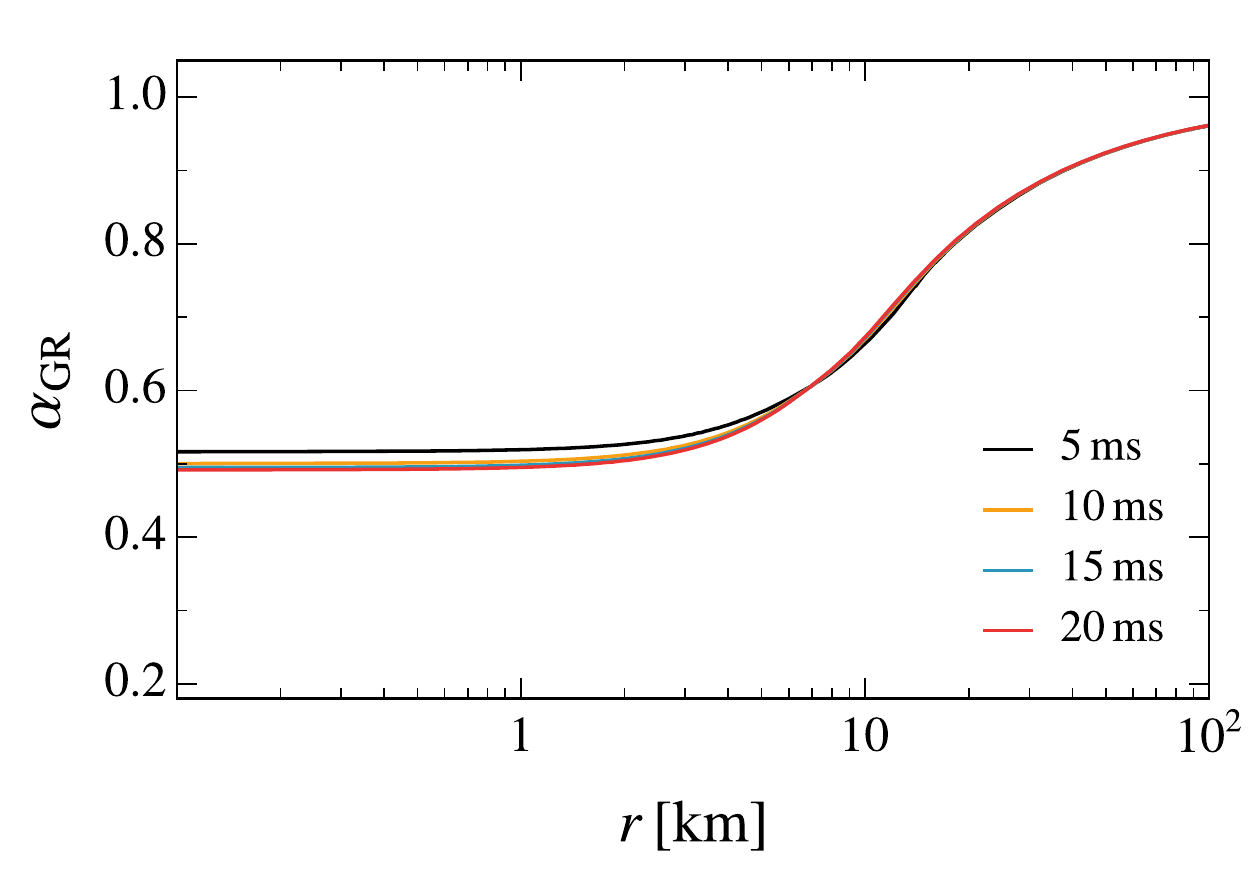}
    \caption{ Radial evolution at different post-merging times of the lapse factor $\alpha_{\rm{GR}}(r)$.}
    \label{fig:lapseface}
\end{figure}
The lapse factor modifies energy and time as follows:
\begin{equation}
\label{eq:alpha}
\begin{split}
    E &= E^*(r)\,\alpha_{\rm{GR}}(r)  \, , \\
    dt &= dt^{*}(r) \,\alpha^{-1}_{\rm{GR}}(r)  \, ,
\end{split}
\end{equation}
where $E^*(r)$ and $dt^*(r)$ are the local energy and time interval, while $E$ and $dt$ are the same quantities as observed at infinity. The relations in Eq.~\eqref{eq:alpha} imply that
\begin{equation}
    d E^{*} d t^{*} = d E d t \, .
\end{equation}
In Fig.~\ref{fig:lapseface} we show the evolution of the radial profile of
$\alpha_{\rm{GR}}(r)$ for different times after merging.  We see that $\alpha_{\rm GR}\sim 0.5$ for $r \lesssim 10$~km, while it approaches 1 at $r\gtrsim 100$~km.

Given the above considerations, starting from the expressions for the ALP production spectrum per unit volume  $d^2n_a/dE_adt$ introduced in Ref.~\cite{Lella:2022uwi}, the red-shifted ALP spectrum is obtained by integrating over the volume of the merger remnant using the angle averaged profiles and time,
\begin{equation}
    \frac{d N}{d E} = \int d V d t \,\frac{d^2 n_a}{dt\,dE} = \int d V d t^* \,\frac{d^2 n_a}{dt^*\,dE^*} \,\alpha^{-1}_{\rm{GR}}(r) \, .
    \label{eq:spectra}
\end{equation}

\begin{figure}[t]
    \centering
    \includegraphics[scale=0.7]{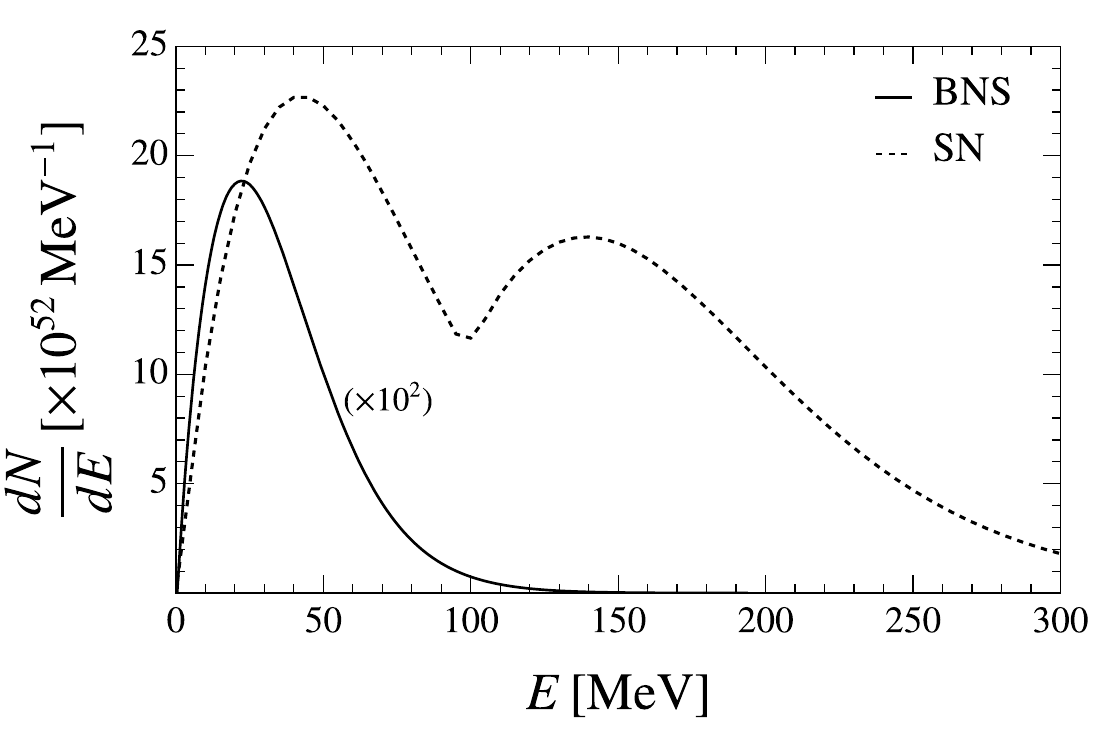}
    \caption{{Time-integrated ALP emission spectrum by means of $NN$  bremsstrahlung from the BNS (solid curve) and from SN (dashed curve). Here we set $g_{a p}=10^{-10}$ and $g_{a n}=0$. Note that in the case of the SN the pionic processes are included.} } 
    \label{fig:dNdE1svs20ms}
\end{figure}

In the time integration, we assume that the input quantities of the BNS models  remain constant from 20~ms up to 1 s after merging, as done in previous ALP studies~\cite{Dev:2023hax,Diamond:2023cto}. {This approach may somewhat overestimate the strength of the signal because the cooling time scale of the merger remnant is of order 1~s and the merger remnant may collapse to a black hole, which would significantly reduce the production of ALPs.} 

The resulting time-integrated spectrum is shown in Fig.~\ref{fig:dNdE1svs20ms} (solid curve), where we set $g_{ap}=10^{-10}$ and $g_{an}=0$.
{ Since $m_a \ll T$, the ALP spectrum can be approximated  with excellent precision using the following analytical expression~\cite{Payez:2014xsa}:
\begin{equation}
    \frac{dN}{dE}=C \left(\frac{g_{a p}}{ 10^{-10}}\right)^{2}\left(\frac{E}{E_0}\right)^\beta \exp\left(-\frac{(\beta+1)E}{E_0}\right)\,,
    \label{eq:alphafit}
\end{equation}
where we obtain $C=1.57 \times 10^{52} \, \text{MeV}^{-1}$, $E_0=36.9\,\text{MeV}$ , and $\beta=1.3$.} 
For comparison, in Fig.~\ref{fig:dNdE1svs20ms} we show as a black dashed curve the SN ALP spectrum obtained by setting $g_{ap}\sim 10^{-10}$ and $g_{an}=0$, in analogy with the BNS case. We highlight that for SN emission the spectrum presents two peaks: the first at $E\sim 50$~MeV associated with the $NN$ process, and the second at $E\sim 150$~MeV  due to the $\pi N$ production.
Comparing the ALP spectrum from BNS mergers with that from SNe, we realize that the first one is peaked at lower energies $E\sim 30$~MeV and it is suppressed by $\sim 2$ orders of magnitude with respect to the SN one for the same ALP-nucleon couplings. This trend can be related to the lower temperature of BNS matter compared to that in the SN core. In addition, the stronger degeneracy of the nucleon population leads to a further reduction of the emission rates. {Finally, ALP emission from BNS mergers has been assumed to occur on a time scale of $\lesssim 1$~s, shorter than the $\sim 10$~s emission from SNe.} All of these effects go in the same direction, resulting in a suppression of the ALP emission rate in BNS mergers with respect to the SN case. Although assuming two identical neutron stars in the merging process typically enhances the temperature of the profile~\cite{Diamond:2023cto}, we note that averaging the thermodynamic properties of the merger remnant may lead to an underestimation of the emission because of the non-linear temperature dependence. Averaged out, high-temperature regions may affect the emissivity more significantly. We also remark that the average temperature of the merger remnant approximately increases with the softness of the EoS~\cite{Kochankovski2025} (see Fig.~18 therein), while we adopted the rather stiff DD2 EoS~\cite{Typel2010} in our merger model.

In the following, when accounting for both ALP-nucleon and ALP-photon couplings, we consider an exemplary scenario inspired by the Kim-Shifman-Vainshtein-Zakharov (KSVZ) axion model~\cite{Kim:1979if,Shifman:1979if}. In this case, the ALP couplings to photons and protons can be related as follows:
\begin{equation}
    g_{ap}=\frac{2\pi\,m_{N}}{\alpha}\frac{C_{ap}}{C_{a\gamma}} g_{a\gamma} \,\,, 
\label{eq:relationcoupling}
\end{equation}
{where $\alpha$ is the electromagnetic fine-structure constant, and $C_{ap}\simeq-0.47$ and $C_{a\gamma}\simeq-1.92$ are model dependent constants set by the UV completion of the model~\cite{GrillidiCortona:2015jxo}. Here, we remark that $g_{an}\simeq0$ in the canonical KSVZ axion model.}
{ For the couplings of this model,  the  bremsstrahlung production rate is always dominant over the Primakoff one associated with $g_{a\gamma}$, so we will neglect this latter hereafter.}
 We also stress that for axion-nucleon couplings below the SN 1987A bound, 
i.e., $g_{aN} \lesssim 10^{-9}$, it was shown in
Ref.~\cite{Harris:2020qim} that ALPs would be in a free-streaming regime, so we can safely neglect reabsorption  in the nuclear medium inside the dense  BNS merger environments.

\section{ALP-photon conversions in magnetic fields}
\label{sec:megnetic} 

\subsection{BNS magnetic field}
\label{sec:BNSmag}
After being emitted during the BNS merger event, ALPs travel long distances before reaching the observer located on the Earth. Notably, along their path from the emission site, ALPs can encounter regions hosting magnetic fields around the remnant, within the host-galaxy, in the intergalactic medium and the Milky Way. 
In particular, the post-merger remnant is expected to be surrounded by a strongly-magnetized medium. The evolution of the BNS merger system is usually described by complex general-relativity magnetohydrodynamics simulations tracking the evolution of the BNS system from the pre- to post-merging phase~(see Refs.~\cite{Paschalidis:2016agf,Baiotti:2016qnr,Duez:2018jaf,Ciolfi:2020cpf} for some recent reviews on the topic).
Initial magnetic fields are significantly amplified during the merging and subsequent evolution of the differentially rotating merger remnant by various processes, such as the Kelvin-Helmholtz instability in the shear layer, magnetic winding, and the development of the magneto-rotational instability. Ultimately, magnetic fields in the remnant and its surroundings can even exceed $10^{15}$~G.
For the sake of simplicity, in the following we assume that a few milliseconds after the merging the remnant $B$-field can be modeled through a dipolar magnetar-like structure,
\begin{equation}
    B(r)=B_{0}\,\left(\frac{r_0}{r}\right)^3\,,
\end{equation}
where $r_0=10$~km is the remnant radius and we conservatively assume a surface magnetic field strength $B_0=10^{15}\,$G. 

For relativistic ALPs following radial trajectories, 
{ the conversions   within the remnant magnetosphere of ALPs into photons}
 can be tracked by employing a Schr\"odinger like equation of the form~\cite{Raffelt:1987im,Fortin:2018ehg,Dessert:2019sgw}
\begin{equation}
    \left(i\frac{d}{dr}-E\right)\begin{pmatrix}
        \gamma_{\parallel}(r)\\
         a (r)\\
    \end{pmatrix}
    =
    \begin{pmatrix}
        \Delta_{\rm HE}^\parallel(r)   &\Delta_{a\gamma}\\
        \Delta_{a\gamma}    &\Delta_a
    \end{pmatrix}
    \,\begin{pmatrix}
        \gamma_{\parallel}(r)\\
        a(r)\\
    \end{pmatrix}\,,
\end{equation}
where $\Delta_{\rm HE}^\parallel$ is the QED vacuum polarization term in the Euler-Heisenberg limit~\cite{Heisenberg:1936nmg}, $\Delta_{a\gamma}$ is the ALP-photon mixing term, and $\Delta_{a}$ is the ALP mass term. We highlight that in this expression the ALP only mixes with the polarization of the photon field parallel to the projection of ${\bf B}$ in the transverse plane with respect to the beam propagation direction ${\bf B}_{T}$. Moreover, in this context we have
\begin{equation}
    \begin{split}
        \Delta^{\parallel}_{\rm HE}&=\,\frac{7\,\alpha}{90\pi B_{\rm crit}^{2}}\,E\,B_T^2\simeq 4.7\times10^{15} \left(\frac{E}{10\,\mathrm{MeV}}\right)\left(\frac{B_0}{10^{15}\,{\rm G}}\right)^2\left(\frac{r}{r_0}\right)^{-6}\,\mathrm{km}^{-1}\,,\\
        \Delta_{a\gamma}&=\frac{1}{2}\,g_{a\gamma}\,B_T\simeq4.9\times10^{3}\,\left(\frac{g_{a\gamma}}{10^{-11}\,\mathrm{GeV}^{-1}}\right)\left(\frac{B_0}{10^{15}\,\mathrm{G}}\right)\left(\frac{r}{r_0}\right)^{-3}{\rm km}^{-1} \,\ ,\\
        \Delta_{a}&=-\frac{m_a^2}{2E}\simeq2.5\times10^{-6}\,\left(\frac{m_a}{10^{-4}\,\eV}\right)^2\left(\frac{E}{10\MeV}\right)^{-1}\,,
        \\     
    \end{split}
    \label{eq:BNSconversion}
\end{equation}
with $B_{\rm crit}=4.41\times10^{13}\,$G. For simplicity, in these expressions we neglect the angular dependence in the remnant magnetic field. From Eq.~\eqref{eq:BNSconversion}, it is apparent that ALP-photon conversions are strongly suppressed by QED vacuum effects close to the remnant surface. Conversely,  we observe that conversion probabilities become sizable at $r\gtrsim10^{4}\,r_0$, where $ \Delta^{\parallel}_{\rm HE}\sim \Delta_{a\gamma}$~\cite{Fortin:2018ehg}. 

\begin{figure}[t]
\centering
    \includegraphics[width=1\textwidth]{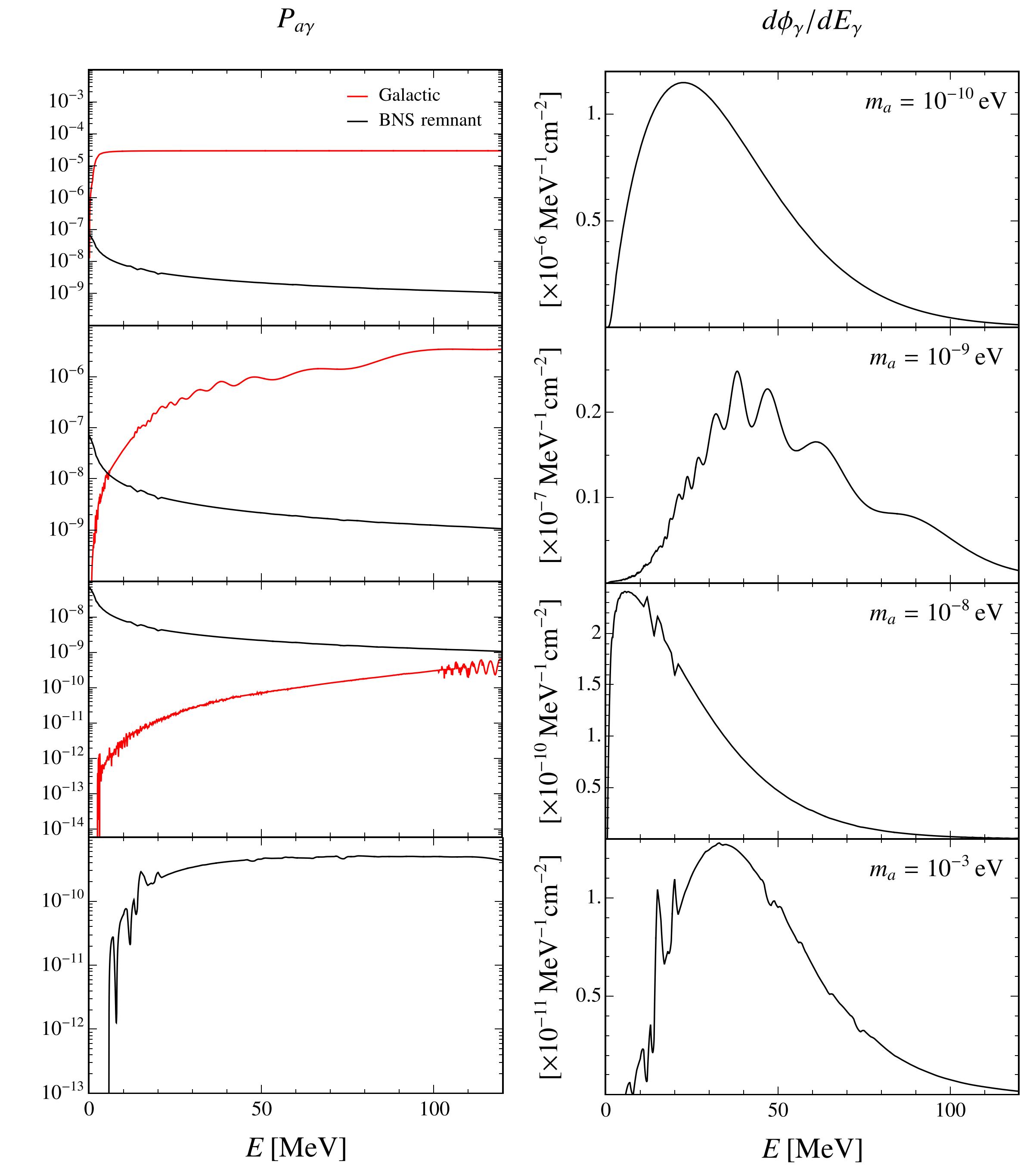}
    \caption{
    \emph{Left panels}: ALP-photon conversion probability as a function of the ALP energy for  $g_{a\gamma}=10^{-12}\,\GeV^{-1}$, in the BNS remnant field (black curves) and the Milky Way magnetic field. Conversions in the Milky Way are computed by modeling the Galactic magnetic field as in Ref.~\cite{Jansson:2012pc} and assuming an ALP beam propagating in the direction $(\ell,b)=(308.38^{\circ}, 39.30^{\circ})$ (red curves). 
    \emph{Right panels:} Photon spectra induced by ALP-photon conversions in sequence in  the BNS remnant magnetic field and the Galactic magnetic field. ALP emission is assumed to occur from a BNS event located at $d=40\,$Mpc from the observer on the Earth. From top to bottom the panels refer to ALPs with masses ${m_a=10^{-10},10^{-9},10^{-8},10^{-3}\,}$ eV, respectively. Note that at $m_a\sim10^{-3}\,\rm{eV}$ the conversion probability in the Milky Way magnetic field is negligible.}
     \label{fig:PhotonFluxes}
\end{figure}

{We plot the conversion probability $P_{a\gamma}$ in the BNS remnant field as the black curves in the left panels of Fig.~\ref{fig:PhotonFluxes}, assuming 
$g_{a\gamma}=10^{-12}\,\GeV^{-1}$.} 
We realize that for $m_a \lesssim 10^{-7}$~eV, $P_{a\gamma}$ is in the range $10^{-7}-10^{-9}$. Then, for ALP masses $m_a\gtrsim10^{-4}\,$eV,   
 at the surface $r\simeq10^{4}\,r_0$ one finds $\Delta_a\gtrsim\Delta_{a\gamma}$. Thus, ALP-photon oscillations become incoherent and energy-dependent. Therefore, as shown in the lower left panel for $m_a=10^{-3}\,$eV, related conversion probabilities are strongly suppressed in this mass range.

\subsection{Milky Way magnetic field}
\label{sec:milky}
As they are extremely weakly interacting, the probability of ALPs escaping the remnant magnetosphere to interact with matter in the interstellar medium is negligible for all practical purposes. 
Furthermore, for the range of parameters we are interested in, the 
ALP lifetime
\begin{equation}
\tau_{a\gamma \gamma}= 1.32 \times  10^{25} 
\left(\frac{\textrm{eV}}{m_a} \right)^3 
\left(\frac{10^{-10}\,\ \textrm{GeV}^{-1}}{g_{a\gamma}}\right)^2 \,\ \textrm{s} \,\, 
\end{equation}
is much larger than the age of the Universe. Moreover, {as discussed in Sec.~\ref{sec:BNSmag},} in the range of couplings $g_{a\gamma}<10^{-10}$, only a negligible fraction of ALPs can be converted into photons in the remnant magnetosphere. Then, we are allowed to assume that the emitted ALP flux during the BNS merger remains unaffected when escaping the remnant volume and can encounter other large-scale astrophysical magnetic fields along its path.

Concerning the galaxy hosting the BNS merger event, one would expect $B$-fields characterized by the same strength of the Milky Way, i.e., ${\mathcal O}(1)$~$\mu$G, on similar size of $\sim 10$~kpc~\cite{Fletcher:2011fn,Beck:2013bxa,2017MNRAS.469.3185P}. As we will comment below, for such values conversion probabilities are expected to be
${\mathcal O}(10^{-4})$. Therefore, one does not expect a sizable depletion of the initial ALP flux due to this effect.  Conversely, the produced photon flux by conversion in the host galaxy would be of the same level as that produced by conversions in the Milky Way. However, lacking detailed models of the magnetic fields,  in order to avoid additional uncertainties we prefer to neglect ALP conversions in the host galaxies. This is a conservative choice that implies that we are neglecting a component of the gamma-ray flux produced by ALP conversions at the same level of the one produced in the Milky Way.
Then, magnetic fields in intergalactic regions may play a significant role in ALP-photon conversions, depending on the parameters characterizing these fields~\cite{Mirizzi:2009aj,Kartavtsev:2016doq}. However, extragalactic magnetic fields are subject to very large uncertainties regarding their strength and typical correlation scales~\cite{Taylor:2011bn, Durrer:2013pga, Caprini:2015gga}. 
Therefore, in our estimation of the ALP conversion probability, we also conservatively neglect the contribution from magnetic fields outside the Milky Way.

On the other hand, the large-scale Galactic magnetic field can be well described by different models able to reproduce related gamma-ray observations~\cite{Pshirkov:2011um,Jansson:2012pc,Unger:2023lob}. In this work, we assume the Jansson-Farrar model~\cite{Jansson:2012pc} as a benchmark model for the Milky Way regular magnetic field, which takes into account a disk field and an extended halo field with an out-of-plane component, based on the WMAP7 Galactic synchrotron emission map~\cite{Gold:2010fm} and extra-galactic Faraday rotation measurements. In particular, we employ the updated parameters given in Table~C.2 of Ref.~\cite{Planck:2016gdp} (“Jansson12c” ordered fields), matching the polarized synchrotron and dust emission measured by the Planck satellite~\cite{Planck:2015mrs,Planck:2015qep,Planck:2015zry}. 

In general, the description of the ALP propagation throughout the Galactic magnetic field model requires a full three-dimensional approach. In this work, we closely follow the treatment outlined in Ref.~\cite{Horns:2012kw} to numerically solve the dynamics of the coupled ALP-photon system along a given line of sight. As an exemplary scenario, in the following we consider the case of an ALP beam propagating from a generic source located at a distance $d=40\,$Mpc from the observer in the same sky location as the GW170817 event $(\ell,b)=(308.38^{\circ}, 39.30^{\circ})$, where $\ell$ and $b$ are the Galactic latitude and longitude, respectively~\cite{DES:2017kbs}. Moreover, we trace ALP-photon conversions within the Milky Way magnetic field for $L=20\,\kpc$, which is the typical size associated with the Galactic disk. The {left panels} of Fig.~\ref{fig:PhotonFluxes} depict in red the behavior of the ALP-to-photon conversion probability as a function of the ALP energy and for four representative ALP masses.


As discussed in Ref.~\cite{Calore:2023srn}, for ALP masses 
\begin{equation} 
   m_a \ll  0.36\,{\rm neV}\left(\frac{E_a}{100 \,\ {\rm MeV}} \right)^{1/2}\left(\frac{L}{10 \,\ {\rm kpc}} \right)^{-1/2}\,\ ,
    \label{eq:mcrit}
\end{equation}
ALP-photon oscillations become energy-independent. In this regime, 
the conversion probability can be well approximated by
\begin{equation}
P_{a\gamma} \simeq 2.3\times10^{-4}\,\left(\frac{g_{a\gamma}}{10^{-12}\,\GeV^{-1}}\right)^2\,\left(\frac{L}{10\,\kpc}\right)^2\,\left(\frac{B_T}{10^{-6}\,\mathrm{G}}\right)^2\,\,,
\label{eq:enindep}
\end{equation}
where $B_T$ is the magnitude of the transverse component of the magnetic field averaged along the given line of sight~\cite{Calore:2023srn}. In this regard, we point out that the Galactic ALP-photon conversion probability depends on the arbitrary choice of the line of sight along which the event is located. To evaluate the impact of this choice on the analysis, we estimate the value of the conversion probability by averaging over all of the possible lines of sight. We obtained that this procedure leads to a variation of less than a factor of $2$ in the conversion probability.
We can observe that for ALPs with mass $m_a = 10^{-10}$~eV the condition in Eq.~(\ref{eq:mcrit}) is always met in the energy range of interest for this study,  and conversions in the BNS remnant are subleading with respect to those in the Milky Way. 
We point out the different trends of the conversion probability at low energies for the Galactic magnetic field compared to the source field. In particular, for $E\rightarrow0$, the mass term, which dominates ALP-photon conversions in the Milky Way, is rapidly increased. Therefore, the Galactic conversion probability becomes energy dependent and gets suppressed. On the other hand, conversions in the remnant field are dominated by QED vacuum effects, which are washed at $E=0$ [see Eq.~\eqref{eq:BNSconversion}]. Thus, ALP-photon oscillations can take place in the inner region of the remnant magnetosphere where fields are stronger, resulting in a larger conversion probability.

In the coherent regime, the induced gamma-ray flux 
\begin{equation}
    \frac{d\phi_\gamma}{dE_\gamma}(E,L)=\frac{1}{4\pi d^2}\,\frac{dN}{dE}(E)\,P_{a\gamma}(E,L) \,\ ,
    \label{eq:photonflux}
\end{equation}
reproduces the same spectral shape of the ALP emission rate from the BNS event~(see right panels of Fig.~\ref{fig:dNdE1svs20ms}). On the other hand, the Galactic conversion probability for ALPs with masses $m_a = 10^{-9}$~eV starts to be suppressed in the energy range of interest, {still remaining dominant with respect to the conversion probability in the remnant magnetic field.} Thus, the related gamma-ray spectrum appears one order of magnitude smaller with respect to the previous case and shows a peculiar oscillatory behavior in the energy range $E\lesssim100\,$MeV.
Finally, ALP-photon oscillations in the Galaxy become incoherent across the entire energy range considered for ALPs with masses $m_a > 10^{-9}$~eV.
As a result, for $m_a > 10^{-8}$~eV, conversions in the BNS remnant become the dominant contribution.
Nevertheless, even when accounting for the magnetic field of the remnant, the gamma-ray flux associated with ALPs of mass $m_a$ larger than a few neV is significantly suppressed.

\section{Sensitivity of gamma-ray experiments to ALPs from BNS mergers}
\label{ref:gammaray}

\subsection{Gamma-ray experiments}
\label{sec:Gamma-ray experiments}

\begin{figure}[!t]
    \centering
    \includegraphics[width=0.7\linewidth]{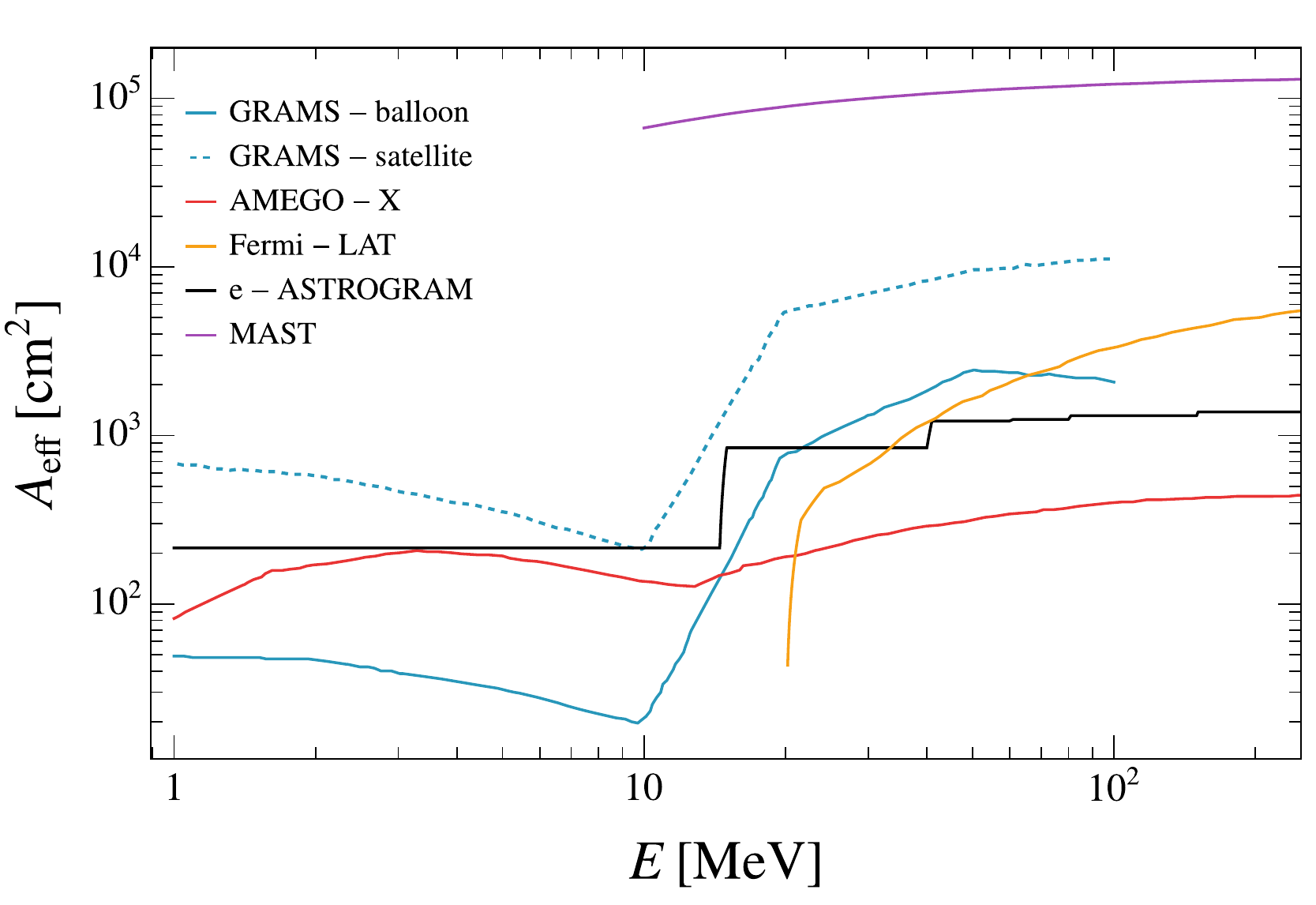}
    \caption{Effective areas of e-ASTROGRAM~\cite{e-ASTROGAM:2016bph}, AMEGO-X~\cite{Caputo:2022xpx}, \emph{Fermi}-LAT~\cite{Fermi-LAT:2021wbg}, GRAMS-balloon~\cite{Aramaki:2019bpi}, GRAMS-satellite~\cite{Aramaki:2019bpi}, and MAST~\cite{Dzhatdoev:2019kay} as functions of the energy of the incoming photon in the energy range relevant for this work.}
    \label{fig:Aeff}
\end{figure}

The detectability of the gamma-ray burst induced by conversions of ALPs emitted during BNS events is directly related to the  properties of current and planned gamma-ray detectors operating in the 1–200 MeV energy ranges.
Among the ones considered in this work, only \emph{Fermi}-LAT is currently operational, while the others have been proposed for future missions. 
The main characteristics of these experiments are summarized  in Fig.~\ref{fig:Aeff} and Table~\ref{tab:exp}. 
In particular, Fig.~\ref{fig:Aeff} displays the effective area of the different experiments in their operational energy ranges.
This feature is crucial in determining whether the expected gamma-ray spectrum falls within the experiment’s sensitivity. We observe that proposed gamma-ray experiments such as e-ASTROGRAM~\cite{e-ASTROGAM:2016bph} and AMEGO-X~\cite{Caputo:2022xpx} would be able to cover the entire range of energies associated with ALP emission from BNS merger events. 
Conversely,  \emph{Fermi}-LAT~\cite{Fermi-LAT:2021wbg} is subject to a dramatic loss of sensitivity at energies $E\lesssim30\,\MeV$, where a significant fraction of the flux is expected. Similarly, both GRAMS-satellite and GRAMS-balloon~\cite{Aramaki:2019bpi} are limited to detecting photons up to $100\,\rm{MeV}$, missing the high-energy tail of the spectrum. 
Therefore, the effects of the large effective areas of Fermi-LAT and GRAMS-satellite on the sensitivity is counterbalanced by the lack of coverage of the entire energy range of interest. 
It is worth noting that GRAMS-satellite offers an improvement in the effective area of approximately one order of magnitude compared to its balloon counterpart. Indeed, according to Ref.~\cite{Aramaki:2019bpi},  
the satellite version, which could be launched in the 2030s, will incorporate upgrades to the detector that are expected to significantly enhance the experiment's sensitivity.
Finally, MAST~\cite{Dzhatdoev:2019kay} exhibits a higher effective area for $E > 10\,\rm{MeV}$. However, since in Ref.~\cite{Dzhatdoev:2019kay} other specifics of the detector were defined only above $100\,\rm{MeV}$, our analysis is conducted assuming this latter 
energy value as a threshold.
From Tab.~\ref{tab:exp}, we observe that GRAMS-satellite and GRAMS-balloon feature a broad field of view, FoV$=6.3$ sr, covering half of the sky for each single pointing. This significantly increases its probability of detecting an ALP-induced gamma-ray signal from a BNS merger event. 


\begin{table}
 \centering
\begin{tabular}{c  c  c  c}
\hline
{ Experiment} & \,\,\,\,\,\,\,\,\,\,\,\,\,\,\, { FoV} \,\,\,\,\,\,\,\,\,\,\,\,\,\,\,& \,\,\,\,\,\,\,\,\,\,\,\,\,\,\, ${\delta \theta}$   
\,\,\,\,\,\,\,\,\,\,\,\,\,\,\,& \,\,\,\,\,\,\,\,\,\,\,\,\,\,\, ${N_{\text{bkg}}}$   
\,\,\,\,\,\,\,\,\,\,\,\,\,\,\, \\ 
 {}  & $(\textrm{sr)} $ & $({}^{\circ})$ & $(\text{counts}\,\,\text{s}^{-1})$ \\
\hline
\hline
e-ASTROGRAM~\cite{e-ASTROGAM:2016bph}  & $\gtrsim 2.5$ & $\lesssim1.5$ & $0.06$ \vspace{0.2cm}\\
AMEGO-X~\cite{Caputo:2022xpx}& $2.5$ & $3$ & $0.25$  \vspace{0.2cm}\\
\emph{Fermi}-LAT~\cite{Fermi-LAT:2021wbg}  & $2.4$ & $\lesssim0.15$ & $0.08$ \vspace{0.2cm}\\
GRAMS-balloon~\cite{Aramaki:2019bpi} & $6.3 $ &  $3$ & $0.27$  \vspace{0.2cm}\\
GRAMS-satellite~\cite{Aramaki:2019bpi}  & $6.3$ & $1.8$ & $0.35$ \vspace{0.2cm}\\
MAST~\cite{Dzhatdoev:2019kay} & $2.5$ & $\lesssim1$ & $0.0004$ \vspace{0.2cm}\\
\hline
\end{tabular}
 \caption{Main features characterizing the current and future gamma-ray experiments considered in this work. 
 }
    \label{tab:exp}
\end{table}

The number of background events $N_{\rm{bkg}}$ for each experiment is shown in Tab.~\ref{tab:exp}. 
For e-ASTROGRAM, this number is obtained by summing the background rate of all of the energy bins of interest provided in Tab.~$4$ of Ref.~\cite{e-ASTROGAM:2016bph}.
For all of the other experiments, the background events are estimated based on a background flux $d\phi_{\gamma,\rm{bkg}}/dE$ expressed in units of $\rm{cm^{-2}\,s^{-1}\,MeV^{-1}\,sr^{-1}}$, assumed to be isotropic.
Given the background flux, the effective area $A_{\rm{eff}}$ illustrated in Fig.~\ref{fig:Aeff}, and the angular resolution of the considered experiment, the number of background events can be computed as
\begin{equation}
  N_{\rm{bkg}} = \Omega \times \int_{E_{\rm min}}^{E_{\rm max}} dE\, \frac{d\phi_{\gamma,\rm{bkg}}}{dE}\, A_{\rm eff}(E)\ ,\, 
  \label{eq:background}
\end{equation}
where 
\begin{equation}
\Omega =2\pi \left(1 - \cos{\delta \theta}\right),
\end{equation}
is the solid angle corresponding to the angular resolution ${\delta \theta}$ (for ${\delta \theta} \ll 1$, 
$\Omega \sim 2 \pi (\delta \theta)^2$), and $E_{\rm min}$ and $E_{\rm max}$ represent the lowest and highest energy values of the considered experiment. 
Specifically, the background events reported in the third column of Tab.~\ref{tab:exp} are obtained using information in Refs.~\cite{Fermi-LAT:2021wbg,Muller:2023pip} for \emph{Fermi}-LAT, Ref.~\cite{e-ASTROGAM:2016bph} for e-ASTROGRAM, Refs.~\cite{Caputo:2022xpx,Carenza:2022som} for AMEGO-X, Ref.~\cite{Aramaki:2019bpi} for GRAMS-satellite and GRAMS-balloon, and Ref.~\cite{Dzhatdoev:2019kay} for MAST.
The specific energy ranges of interest in our study correspond to the energy coverage of the effective area shown in Fig.~\ref{fig:Aeff}.
Among the considered experiments, MAST exhibits the lowest number of background events (see Tab.~\ref{tab:exp}). Conversely, GRAMS-balloon and AMEGO-X, due to their poor angular resolution, yield a higher number of background events. Finally, GRAMS-satellite records the highest number of background events, surpassing even the standard version of the experiment. This increase is attributed to the larger effective area (see Fig.~\ref{fig:Aeff}), which counterbalances the improvements in angular resolution.

\subsection{Sensitivities}
The number of gamma-ray events from  ALPs emitted in BNS merger events  and converting in the  magnetic fields can be estimated as

\begin{equation}
    N_{\rm{ev}} = \int_{E_{\rm min}}^{E_{\rm max}}dE\,\frac{d\phi_{\gamma}}{dE}\, A_{\rm eff}(E) \, ,
\end{equation}

{where ${d\phi_{\gamma}}/{dE}$ is the flux of photons reaching the Earth [see Eq.~(\ref{eq:photonflux})].}
The sensitivities to the ALP-photon coupling for the considered experiments are obtained by requiring that the number of ALP-induced gamma-ray events, $N_{\rm ev}$,  exceeds the number of background events, $N_{\rm{ev}}\,\gtrsim N_{\rm{bkg}} \,$
[see Eq.~(\ref{eq:background})]. 



\begin{table}[t!]
 \centering
\begin{tabular}{c  c  c  c}
\hline
{Experiment} & \,\,\,\,\,\,\,\,\,\,\,\,\,\,\,\,\,\,\,\,\,\,\,\,\,\,\,\,\,\,\,\,\,\,\,\,\,\,\,\,\,\,\,\,\,\,\,\,\,\,\,\,\,\,\,\,\,\,\,\,\,\,\,{} & { Sensitivity on $g_{a\gamma}\,[\times10^{-12}\,{\rm GeV}^{-1}]$}  &  {}\\ 
{} & { $d=4\,\text{Mpc}$}  &  { $d=40\,\text{Mpc}$}  & { $d=100\,\text{Mpc}$} \\ 
 
\hline
\hline
e-ASTROGRAM~\cite{e-ASTROGAM:2016bph} & $0.3$  &  $1.1$ & $1.7$  \vspace{0.2cm}\\
AMEGO-X~\cite{Caputo:2022xpx} & $0.6$ &  $2.0$ & $3.2$  \vspace{0.2cm}\\
\emph{Fermi}-LAT~\cite{Fermi-LAT:2021wbg} & $0.3$ &  $1.1$ & $1.7$  \vspace{0.2cm}\\
GRAMS-balloon~\cite{Aramaki:2019bpi} & $0.4$ & $1.4$ & $2.2$  \vspace{0.2cm}\\
GRAMS-satellite~\cite{Aramaki:2019bpi}  & $0.3$ &  $1.0$ & $1.6$  \vspace{0.2cm}\\
MAST~\cite{Dzhatdoev:2019kay} & $0.08$ &  $0.27$ & $0.4$  \vspace{0.2cm}\\
\hline
\end{tabular}
 \caption{Sensitivities of e-ASTROGRAM~\cite{e-ASTROGAM:2016bph}, AMEGO-X~\cite{Caputo:2022xpx}, \emph{Fermi}-LAT~\cite{Fermi-LAT:2021wbg}, GRAMS-balloon~\cite{Aramaki:2019bpi}, GRAMS-satellite~\cite{Aramaki:2019bpi}, and MAST~\cite{Dzhatdoev:2019kay} in the massless ALP limit for a signal coming from a source located at $d = 4, 40, 100\,\mathrm{Mpc}$ in the same direction as the GW$170817$ event, { assuming $g_{a\gamma}$ and $g_{ap}$ related as in Eq.~\eqref{eq:relationcoupling}}.}
    \label{tab:sensitivitiesBrem}
\end{table}


Table~\ref{tab:sensitivitiesBrem}  shows the sensitivities of the six experiments in the massless ALP limit ($m_{a}\lesssim 10^{-10}\,$eV) for a signal coming from a source located at $d = 4, 40, 100 \,\mathrm{Mpc}$ in the same direction as the GW$170817$ event. We recall that, on average, choosing another sky location would affect the conversion probability by a factor of approximately a few (see Sec.~\ref{sec:milky}). Nevertheless, since the photon fluxes scale as $\sim g_{a\gamma}^4$, our limits on the ALP-photon coupling go as $\sim P_{a\gamma}^{1/4}$. Thus, our limits are expected to show a weak dependence on the sky location of the given event. Here $g_{ap}$ and $g_{a\gamma}$ are correlated as in Eq.~\eqref{eq:relationcoupling}.
{We find that the best sensitivity is achieved by MAST, which probes $g_{a\gamma}\gtrsim 0.27\times 10^{-12}\,\text{GeV}^{-1}$
for $d = 40\,\mathrm{Mpc}$, while at that distance the other experiments have similar sensitivities, $g_{a\gamma}\gtrsim 10^{-12}\,\text{GeV}^{-1}$.}

\begin{figure}
    \centering
    \includegraphics[width=0.7\linewidth]{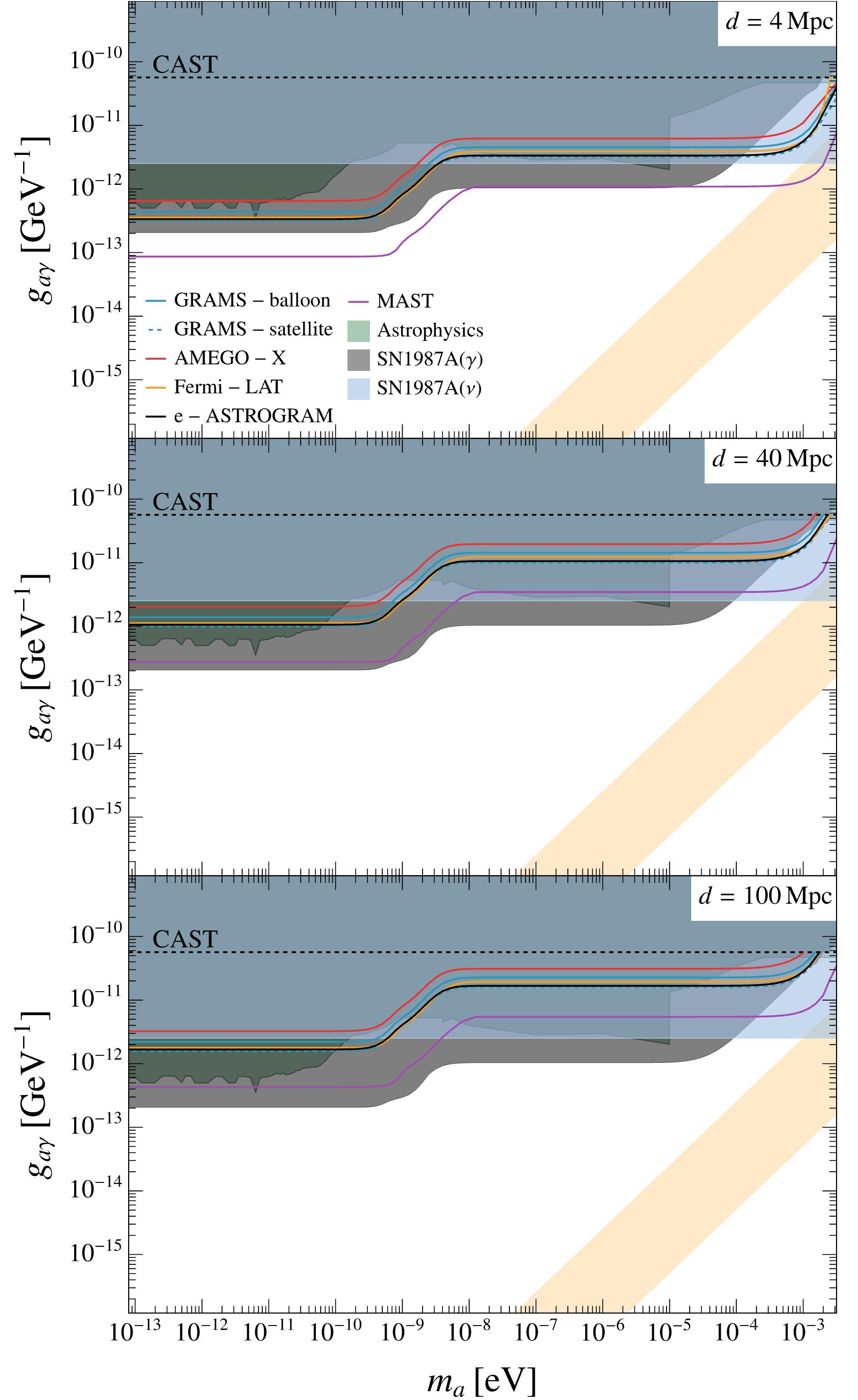}
    \caption{Sensitivities of e-ASTROGRAM~\cite{e-ASTROGAM:2016bph}, AMEGO-X~\cite{Caputo:2022xpx}, \emph{Fermi}-LAT~\cite{Fermi-LAT:2021wbg},  GRAMS-balloon~\cite{Aramaki:2019bpi}, GRAMS-satellite~\cite{Aramaki:2019bpi}, and MAST~\cite{Dzhatdoev:2019kay} in the plane $g_{a\gamma}$ \emph{vs} $m_a$ assuming that
    $g_{ap}$ and $g_{a\gamma}$ are related as in Eq.~\eqref{eq:relationcoupling}.
    We assume a {BNS system at the same sky location as} the GW$170817$ event, at a distance of $d = 4\,\text{Mpc}$ (upper panel), $d = 40\,\text{Mpc}$ (middle panel), and $d=100\,\text{Mpc}$ (lower panel) from the observer, respectively. The gray bound represents the limit for SN 1987A derived  {using the spectra shown in Fig.~\ref{fig:dNdE1svs20ms} and taking into account the conversion probability in the progenitor and the Milky Way magnetic fields}~\cite{Payez:2014xsa,Hoof:2022xbe}, and the dashed line shows the most recent CAST bound from solar axions~\cite{CAST:2024eil}. Finally, the bounds in green correspond to constraints from other astrophysical observations~\cite{Ning:2024eky,Fermi-LAT:2016nkz,Li:2024zst,Noordhuis:2022ljw}. We highlight that bounds in green are derived by assuming ALPs coupled to photons only.}
    \label{fig:bound+sensGWbremres}
\end{figure}

In Fig.~\ref{fig:bound+sensGWbremres}, we show 
the sensitivities associated with the different experiments in the plane $g_{a\gamma}$ \emph{vs}
$m_a$ 
for a BNS system at the same sky location as the GW$170817$ event
at $d=4$~Mpc (upper panel),  $d=40$~Mpc (middle panel), and
$d=100$~Mpc (lower panel).
Just like we did in Tab.~\ref{tab:sensitivitiesBrem}, also here we also assume that $g_{ap}$ and $g_{a\gamma}$ are correlated as in Eq.~\eqref{eq:relationcoupling}.
The shapes of these sensitivity curves exhibit  a plateau in $g_{a\gamma}$  for $m_a \lesssim 10^{-9}~\eV$ that corresponds to coherent ALP-photon conversions in the magnetic field of the Milky Way. For larger masses, the sensitivity deteriorates, until conversions in the magnetic field of the BNS mergers become dominant in the mass range $10^{-8}~\eV$ $\lesssim m_a \lesssim$ $10^{-3}~\eV$, producing another plateau. For higher values of  $ m_a$, the conversions in the remnant are also strongly suppressed, producing a strong worsening in the sensitivity on $g_{a\gamma}$  (see Fig.~\ref{fig:PhotonFluxes}).

The green region in Fig.~\ref{fig:bound+sensGWbremres} is excluded by astrophysical constraints~\cite{Fermi-LAT:2016nkz,Noordhuis:2022ljw,Li:2024zst,Ning:2024eky,Manzari:2024jns}.  For comparison, we also show {the most recent CAST  bound from solar ALPs} (dashed black curve)~\cite{CAST:2024eil}. {For the sake of clarity, we highlight that these constraints are placed by considering ALPs coupling to photons only.}
We also depict in yellow the band where preferred QCD axion models live (see, e.g., Ref.~\cite{Carenza:2024ehj}).
We remark that the BNS sensitivities as the other represented bounds (except the CAST one at higher masses) are very far to touch QCD axion models.
An exception is the case of an ALP signal from a very nearby BNS merger at $d=4$~Mpc, where the sensitivity of the MAST experiment would touch the QCD axion band for $m_a \sim 10^{-3}~\eV$.
{ Furthermore, we show in light blue the bound from SN 1987A based on an excessive shortening of the observed neutrino burst, in the presence of a copious ALP emission by nuclear processes~\cite{Lella:2023bfb}.}
{{In order to have a direct comparison with a similar physics case,} we also update the long-standing bound from the non-observation of a gamma-ray signal in coincidence with SN 1987A~\cite{Payez:2014xsa,Hoof:2022xbe}, taking into account 
as the initial ALP spectrum the one shown in Fig.~\ref{fig:dNdE1svs20ms}.
Furthermore, we characterize 
 ALP-photon conversions in the  magnetic field of the SN, following the same model of Ref.~\cite{Manzari:2024jns},
 while for conversions in the Milky Way we refer to
Sec.~\ref{sec:milky}.
 We remark that the conversions in the SN magnetic field produce a plateau in the exclusion plot in the mass range $10^{-8}~\eV$ $\lesssim m_a \lesssim$ $10^{-4}~\eV$, due to an enhancement in the conversions for higher ALP masses, associated with the strong $B$-field in the source. }
 {Notice that the factor $\sim 2$ difference with respect to the results of Ref.~\cite{Manzari:2024jns} has to be attributed to different models of SN and Galactic magnetic fields, as well as the analysis methodologies adopted in the two works.} 
{We realize that, due to the lower emissivity of ALPs from BNS mergers, the SN 1987A bound is always stronger than possible sensitivities from BNS mergers, expect for the case of the MAST experiment. Indeed, due to an extremely small background, MAST can probe ALP-photon couplings comparable to the SN 1987A one for ALP masses $m_a \lesssim 10^{-9}~\eV$ and $d=40$~Mpc, and even better in the best-case scenario
of  $d=4$~Mpc.}

\subsection{Probability of joint GW-gamma detection}

Finally, a key aspect to clarify is the typical rate at which a gamma-ray signal from extragalactic BNS mergers might be expected.
This rate depends on the probability of observing an extra-galactic ALP-induced gamma-ray burst simultaneously with GW detection.
Therefore, first one should evaluate the rate at which GWs from BNS systems can be observed  by GW detectors.
The authors of Ref.~\cite{Pol:2018shd} proposed using 
the $B$-luminosity, i.e., the luminosity in the blue band, as a good tracer of recent star formation in the Universe. Therefore, 
starting with the estimated rate of BNS mergers in the Milky Way, one can extrapolate it to extra-galactic distances, rescaling it  through the ratio of 
$B$-luminosity at a certain distance $d$ with respect to that in the Galaxy {$L_{\rm MW}$}. 
Specifically, one gets~\cite{Pol:2018shd}
\begin{equation}
\mathcal{R}_{\rm GW}=\mathcal{R}_{\rm{MW}}\left(\frac{L_{\rm{total}}(d)}{L_{\rm{MW}}}\right) \,\ ,
\end{equation}
where the BNS merger rate  in the Milky Way is given by 
\begin{equation}
\mathcal{R_{\rm{MW}}}=42^{+30}_{-14}\,\text{Myr}^{-1}\,\ ,
\label{eq:rateMW}
\end{equation}
as inferred from the Galactic pulsar population.

Choosing $100\,$Mpc as the GW detector horizon,  as in the case of Advanced LIGO~\cite{KAGRA:2013rdx}, 
the resulting GW detection rate would be given by~\cite{Pol:2018shd}
\begin{equation}
R_\text{GW} \sim 0.18^{+0.13}_{-0.06} \times \left( \frac{d}{100 \, \text{Mpc}} \right)^3 \, \text{yr}^{-1}.
\label{eq:rateLIGO}
\end{equation}
Thus, Advanced LIGO is expected to detect a BNS merger event at $100 \, \text{Mpc}$ from Earth approximately once every $3$--$8$ years. We note, however, that for the next observing run starting in a few years, a detection horizon of $200\,$Mpc is anticipated, which would enhance the detection rate by approximately an order of magnitude. 

Based on the above estimate [Eq.~\eqref{eq:rateLIGO}],
one can evaluate 
the probability of a joint gamma-GW signal detection.
The gamma-ray detection probability {$P_{\rm tot}$} is determined by the product of two factors, namely, the probability of the gamma-ray experiment being active during the occurrence of the BNS event {($P_{\rm on}$)}, times the probability of the event falling within the experiment FoV {($P_{\rm FoV}$)}. For all of the experiments, the first contribution is evaluated by assuming a survey mode similar to that of \emph{Fermi}-LAT, i.e., ${P}_{\rm on}=85~\%$, which accounts for the turning off of the instrument when passing through the South Atlantic Anomaly~\cite{Pillera:2024jro}.
Moreover, by assuming isotropic distribution  mergers, the  probability of getting a BNS merger event in the experiment FoV is given by ${P}_{\rm FoV} = {\rm{FoV}}/{4\pi}$.
We remark  that, in the case of MAST, in the absence of  detailed information,  the FoV is assumed to be the same as that of Fermi-LAT.
\begin{table}[t!]
 \centering
\begin{tabular}{c  c  c  c }
\hline
{ Experiment} & \,\,\,\,\,\,\,\,\,\,\,\,\,\,\,${P_{\rm{on}}} $ \,\,\,\,\,\,\,\,\,\,\,\,\,\,\,& \,\,\,\,\,\,\,\,\,\,\,\,\,\,\, ${P_{\rm{FoV}}} $ \,\,\,\,\,\,\,\,\,\,\,\,\,\,\, & \,\,\,\,\,\,\,\,\,\,\,\,\,\,\, ${P_{\rm{tot}}} $ \,\,\,\,\,\,\,\,\,\,\,\,\,\,\,\\ 
\hline
\hline
\emph{Fermi}-LAT, e-ASTROGRAM, AMEGO-X, MAST &  $85\%$ & $19\%$ & $16\%$ \vspace{0.2cm}\\
GRAMS-balloon, GRAMS-satellite &  $85\%$ & $50\%$ & $43\%$ \vspace{0.2cm}\\

\hline
\end{tabular}
 \caption{Total probabilities $P_{\rm tot}$ to observe the portion of the sky in which the BNS event occurs for e-ASTROGRAM~\cite{e-ASTROGAM:2016bph}, AMEGO-X~\cite{Caputo:2022xpx}, \emph{Fermi}-LAT~\cite{Fermi-LAT:2021wbg}, GRAMS-balloon~\cite{Aramaki:2019bpi}, GRAMS-satellite~\cite{Aramaki:2019bpi}, and MAST~\cite{Dzhatdoev:2019kay}.
 In particular, ${P_{\rm on}}$ represents the probability of the experiment being switched on during the occurrence of the BNS event and ${P_{\rm{FoV}}}$ denotes the probability of the event lying within the experiment FoV. With these definitions, $P_{\rm tot}=P_{\rm on}\times P_{\rm FoV}$.
}

    \label{tab:probabilities}
\end{table}
\begin{table}[t!]
 \centering
\begin{tabular}{c  c  c  c}
\hline
{ Experiment} & \,\,\,\,\,\,\,\,\,\,\,\,\,\,\,\,\,\,\,\,\,\,\,\,\,\,\,\,\,\,\,\,\,\,\,\,\,\,\,\,\,\,\,\,\,\,\,\,\,\,\,\,\,{}  \,\,\,\,\,\,\,\,\,\,\,\,\,\,\,\,& \,\,\,\,\,\,\,\,\,\,\,\,\,\,\, { ${T_{\rm{joint}}}$ }\,\,\,\,\,\,\,\,\,\,\,\,\,\,\, & \,\,\,\,\,\,\,\,\,\,\,\,\,\,\,\,\,\,\,\,\,\,\,\,\,\,\,\,\,\,\,\,\,\,\,\,\,\,\,\,\,\,\,\,\,\,\,\,\,\,\,\,\, {}\,\,\,\,\,\,\,\,\,\,\,\,\,\,\, \\ 
 & {  $d=4\,\text{Mpc}$} & {  $d=40\,\text{Mpc}$} &  {  $d=100\,\text{Mpc}$} \\ 
\hline
\hline
\emph{Fermi}-LAT, e-ASTROGRAM, AMEGO-X, MAST &  $\sim (3$--$8) \times10^{5}\,\,$yr & $\sim(3 $--$8) \times10^{2}\,\,$yr & $\sim 20$--$50\,\,$yr \vspace{0.2cm}\\
GRAMS-balloon, GRAMS-satellite  &   $\sim (1 $--$3) \times10^{5}\,\,$yr & $\sim (1$--$3)\times10^{2}\,\,$yr & $\sim 8$--$20\,\,$yr \vspace{0.2cm}\\

\hline
\end{tabular}
 \caption{Average time interval ${T_{\rm{joint}}} $ between two joint GW-gamma detection by employing Advanced LIGO together with e-ASTROGRAM~\cite{e-ASTROGAM:2016bph}, AMEGO-X~\cite{Caputo:2022xpx}, \emph{Fermi}-LAT~\cite{Fermi-LAT:2021wbg}, GRAMS-balloon~\cite{Aramaki:2019bpi}, GRAMS-satellite~\cite{Aramaki:2019bpi}, and MAST~\cite{Dzhatdoev:2019kay} for a BNS merger event located at $d=4, 40, 100\,\text{Mpc}$.
}
    \label{tab:timeinterval}
\end{table}
These probabilities computed for each of the considered experiments are shown in Tab.~\ref{tab:probabilities}. 
Finally, one can estimate the time interval between two joint-detection events by Advanced LIGO and gamma-ray experiments as
\begin{equation}
    T_{\rm joint}\simeq (R_{\rm{GW}}\times {P}_{\rm{on}}\times {P}_{\rm{FoV}})^{-1} \,\ .
\end{equation}
{We show the computed values of $T_{\rm joint}$ for the different considered experiments in Tab.~\ref{tab:timeinterval}, where the given range is attributed to the uncertainty related to the GW detection rate.} 
From Tab.~\ref{tab:timeinterval} we can see that for a source at a distance $d=4\,$Mpc, the expected time to observe a BNS merger event is 
$\sim 10^5$~years, making this event extremely unlikely. Also the probability to get a joint detection at $d=40$~Mpc, i.e., the same distance as GW170817, is less than one per century. 
For sources at a distance of $100\,$Mpc, GRAMS-balloon and GRAMS-satellite are expected to achieve a joint detection of a BNS merger event approximately every $8$--$20$ years. In contrast, the other experiments are expected to achieve joint detections roughly once every $20$--$50$ years. These results highlight the importance of achieving wider sky coverage~\cite{Manzari:2024jns}.
If one assumes the employment of three experiments (for instance, AMEGO-X, \emph{Fermi}-LAT, and GRAMS-satellite) in orbit at different points of the sky at the same time, it would be possible to cover up to $88\%$ of the sky. 
Moreover, the probability of operational downtime could be considered negligible, as at least two out of the three experiments would remain  operational. Under such conditions, the joint detection of a BNS event at $100\,\text{Mpc}$ would be reduced to approximately once every $4$--$9$ years.

{ Finally,  we  remark that a significant improvement in our analysis would be achieved thanks to the third-generation GW detectors, like the Einstein Telescope~\cite{ET:2019dnz} or Cosmic Explorer detector~\cite{Evans:2021gyd}, which could be operating in the mid 2030s with an improvement of one order of magnitude in GW sensitivities from BNS mergers with respect to detectors like Advanced LIGO.
It has been estimated that such  detectors can collect GW signals 
 a few days before the neutron stars merge, for a distance $d\lesssim 200$~Mpc. This opportunity would be useful to get a localization and early warning of a BNS merger event, which could be exploited to allow the gamma-ray detectors to point in advance at the BNS merger and search for the ALP-induced gamma-ray burst. With this exciting  possibility, 
 one would expect to collect at least a BNS merger per decade in the relevant horizon for ALP searches. 
 }

\section{Conclusions}
\label{sec:conclu}

In this work, we have considered the physics potential of a multi-messenger detection of a gamma-ray burst induced by ALPs produced in 
extra-galactic BNS merger events, and then converting into photons in the remnant and Galactic magnetic field.
We used as an external trigger for such an event the detection of the associated GW signal.
We have shown that, {assuming ALPs coupled to both nucleons and photons as in a canonical KSVZ model}, current and planned gamma-ray telescopes operative in the MeV region can reach a sensitivity to the ALP-photon coupling down to
$g_{a\gamma}\gtrsim \textrm{few} \times 10^{-13}\,\text{GeV}^{-1}$ for
$m_a \lesssim 10^{-9}$~eV, comparable with the SN 1987A limit.
We remark that our analysis relies on conservative assumptions, such as spherical averaging over hydrodynamical quantities and ALPs coupling with only protons, which may underestimate the ALP flux and thus the predicted sensitivities.
Furthermore, we have shown that in the most optimistic situation one can hope for a joint GW-gamma-ray signal every 5-10 years in a radius of 100 Mpc. Therefore, given sufficient time, one can accumulate statistics, improving the expected sensitivity through a stacked analysis.

In conclusion, our work shows that multi-messenger BNS merger detection could be exploited to search for ALPs, combining the unprecedented sensitivity of GW interferometers and gamma-ray detectors, to catch signals from  merging neutron stars in the farthest regions of the Universe.

\acknowledgements

We warmly thank Damiano Fiorillo and Edoardo Vitagliano for useful comments on the manuscript. We thank Tobias Fischer for useful discussions and providing analysis scripts. This article is based on work from COST Action COSMIC WISPers CA21106, supported by COST (European Cooperation in Science and Technology).
The work of AM and AL and FL was partially supported by the research grant number 2022E2J4RK "PANTHEON: Perspectives in Astroparticle and
Neutrino THEory with Old and New messengers" under the program PRIN 2022 (Mission 4, Component 1,
CUP I53D23001110006) funded by the Italian Ministero dell'Universit\`a e della Ricerca (MUR) and by the European Union – Next Generation EU. GL acknowledges support from the U.S. Department of Energy under contract number DE-AC02-76SF00515. This work is (partially) supported by ICSC – Centro Nazionale di Ricerca in High Performance Computing. VV and AB acknowledge support by the European Union through ERC Synergy Grant HeavyMetal no. 101071865. AB acknowledges support by the Deutsche Forschungsgemeinschaft (DFG, German Research Foundation) through Project - ID 279384907 – SFB 1245 (subprojects B07).
MG acknowledges support from the Spanish Agencia Estatal de Investigación under grant PID2019-108122GB-C31, funded by MCIN/AEI/10.13039/501100011033, and from the “European Union NextGenerationEU/PRTR” (Planes complementarios, Programa de Astrofísica y Física de Altas Energías). He also acknowledges support from grant PGC2022-126078NB-C21, “Aún más allá de los modelos estándar,” funded by MCIN/AEI/10.13039/501100011033 and “ERDF A way of making Europe.” Additionally, MG acknowledges funding from the European Union’s Horizon 2020 research and innovation programme under the European Research Council (ERC) grant agreement ERC-2017-AdG788781 (IAXO+).

\setcounter{equation}{0}
\setcounter{figure}{0}
\setcounter{table}{0}
\setcounter{section}{0}
\makeatletter
\renewcommand{\theequation}{A\arabic{equation}}
\renewcommand{\thefigure}{A\arabic{figure}}
\renewcommand{\thetable}{A\arabic{table}}

\appendix

\bibliographystyle{bibi.bst}
\bibliography{references.bib}

\end{document}